# Optical Characterization of Epsilon-Near-Zero, Epsilon-Near-Pole and Hyperbolic Response in Nanowire Metamaterials

R. STARKO-BOWES,[1,*] J. ATKINSON,[1,*] W. NEWMAN,[1] H. HU,[1] T. KALLOS,[2] G. PALIKARAS,[2] R. FEDOSEJEVS,[1] S. PRAMANIK,[1] AND Z. JACOB,[1]

1 Department of Electrical and Computer Engineering, University of Alberta, Edmonton, AB T6G2V4, Canada
2 Lambda Guard, Technology Innovation Centre, 1 Research Drive Dartmouth, NS B2Y 4M9, Canada
*R. Starko-Bowes and J. Atkinson contributed equally to this work.
Corresponding author: zjacob@ualberta.ca

We report on the optical and physical characterization of metallic nanowire (NW) metamaterials fabricated by electrodeposition of ≈ 30 nm diameter gold nanowires in nano-porous anodic aluminum oxide. We observe a uniaxial anisotropic dielectric response for the NW metamaterials that displays both epsilon-near-zero (ENZ) and epsilon-near-pole (ENP) resonances. We show that a fundamental difference in the behavior of NW-metamaterials from metal-dielectric multilayer (ML) metamaterials is the differing directions of the epsilon-near-zero (ENZ) and epsilon-near-pole (ENP) dielectric responses relative to the optical axis of the effective dielectric tensor. In contrast to multilayer metamaterials, nanowire metamaterials exhibit an omnidirectional ENP and an angularly dependent ENZ. Also in contrast to ML metamaterials, the NW metamaterials exhibit ENP and ENZ resonances that are highly absorptive and can be effectively excited from free space. Our fabrication allows a large tunability of the epsilon-near-zero resonance in the visible and near IR spectrum from 583 nm to 805 nm as the gold nanorod fill fraction changes from 26% to 10.5%. We support our fabrication process flow at each step with rigorous physical and optical characterization. Energy dispersive X-ray (EDX) and X-ray diffraction (XRD) analyses are used to ascertain the quality of electrochemically deposited Au nanowires prior to and after annealing. Our experimental results are in agreement with simulations of the periodic plasmonic crystal and also analytical calculations in the effective medium metamaterial limit. We also experimentally characterize the role of spatial dispersion at the ENZ resonance and show that the effect does not occur for the ENP resonance. The application of these materials to the fields of biosensing, quantum optics and thermal devices shows considerable promise.

## 1. Introduction

Nanowire (NW) and multilayer (ML) metamaterials made of plasmonic metals are the two most promising designs for device applications that show tunable epsilon-near-zero (ENZ) [1], epsilon-near-pole (ENP) [2] and hyperbolic responses [3]–[5]. These three exotic electromagnetic responses arise from the coupled plasmon-polaritons in the multilayer super-lattice and nanowire geometries. Although both the NW and ML metamaterials can be characterized by an effective uniaxial dielectric tensor the key characteristics of the types of plasmons supported by each of these two structures causes the two materials to display very distinct electromagnetic responses (see Fig. 1). While ML metamaterials are generally easier to fabricate, NW metamaterials offer significant benefits in terms of higher transmission, lower loss and a broad tunability across the visible frequency range [1], [3], [6], [7]. As a result, NW metamaterials are expected to have dramatic impacts in fields such as super-resolution imaging [8], polarization control/filtering [9], thermal photovoltaics (TPV) [2], [10]–[13], and biosensing [6].

Several procedures have been used by different groups to achieve metamaterials of metallic nanowires in a dielectric matrix. The most common and by far the easiest way to fabricate nanowire arrays is to deposit metal via electrodeposition into the pores of anodic aluminum oxide (AAO) templates. These templates are commercially available [4] however additional control over the template properties can be gained by fabricating them in house on either Al foil substrates [14]–[16] or Al thin films that have been deposited on to transparent substrates [3]. Depending on sample constraints, metallic nanowires (typically Au or Ag) can be deposited by either AC [14] or DC [3], [4], [15], [16] electrochemical deposition. A simple yet effective technique to characterize and confirm the ENZ, ENP and hyperbolic responses of the systems is to study spectrally resolved angular transmission measurements [14], [17]. More sophisticated characterization techniques such as Brewster angle discontinuity reflection measurements [4], shifting of Fabry-

Perot oscillations [15] and near-field scanning optical microscopy (NSOM) [16] allow for a deeper understanding of the optical response of the nanowire metamaterials.

In this paper, we report on the optical and physical characterization of nanowire metamaterials emphasizing the key differences between ML and NW metamaterials. Even though a body of research exists on these two nanostructured media, we show that the difference in resonant characteristics arises from the opposite orientations of the ENZ and ENP responses. Our work also shows for the first time the role of the ENP metamaterial and its omnidirectional nature which was recently predicted to have applications in Salisbury screens [18]. For completeness, we present the in depth fabrication procedure of an array of vertically aligned gold nanowires housed in an AAO matrix supported by an optically transparent glass substrate. To optimize the quality of our NW materials we have analyzed the fabrication with rigorous material characterization and structural analysis using scanning electron microscopy (SEM), energy dispersive X-ray (EDX) and X-ray diffraction (XRD) characterization techniques at each step of the process flow. Reliable estimates of NW fill fractions are extracted using post-processing of SEM micrograph images. The tunability of the ENZ and ENP absorption resonances of the nanowire metamaterials were observed using white light transmittance measurements and are presented as extinction spectra for metamaterials containing different fill fraction/nanowire diameters. Theoretical analysis and modelling in the form of Effective Medium Theory (EMT) and full wave numerical simulation (commercial software CST) [19] shows excellent agreement with experimental data. Finally, we experimentally contrast the spatially dispersive properties of the ENZ and ENP resonances. This work will help researchers adopt our approach to large area nanowire fabrication and characterization for possible quantum, thermal and imaging applications [20]–[23].

## 2. Theory

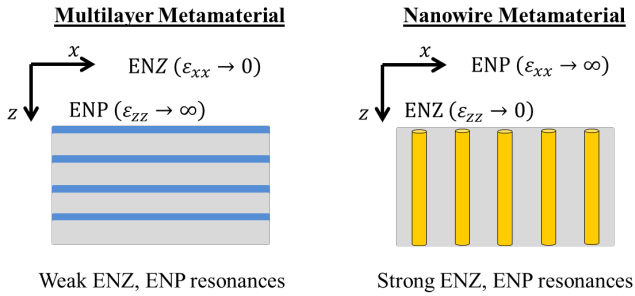

**Figure 1:** Schematic depiction of multilayer vs nanowire metamaterials cross section and the orientation of the ENZ and ENP directions. The difference in directions causes a stark contrast in resonant properties.

ML metamaterials have a periodic structure of alternating sub-wavelength metal and dielectric layers in one dimension ($\vec{z}$). In contrast, NW metamaterials have a periodic structure in two dimensions ($\vec{x} \& \vec{y}$). Both structure types are characterized by an effective dielectric tensor that is uniaxial anisotropic that displays ENZ, ENP and hyperbolic dispersion. However, the directions in which the ENZ and ENP occur are opposite for two structures. Interestingly the NW metamaterials exhibit an angularly dependent ENZ resonance whose spectral location is highly tunable and can be controlled by adjusting the metal nanowire fill fraction. Even though ML metamaterials also exhibit ENP and ENZ resonances as the NW metamaterials do, their resonance characteristics are significantly weaker due to free electron motion in the $\vec{x} \& \vec{y}$ directions leading to high reflection of incoming propagating fields [2], [12], [24].

A useful metric to quantify the absorption spectra of the metamaterials is the extinction (optical density $OD$) spectrum. In this study, the extinction spectrum is defined as $OD = -log_{10}(T)$ where $T$ is the transmittance through the sample. This definition is used theoretically in simulations as well as experimentally to characterize absorbing plasmonic resonances. Peaks will occur in extinction spectra at locations of low transmission. Diffuse reflection measurements indicate that reflection and scattering is relatively weak for our NW samples in the visible and near IR spectrum. This implies that peaks in the extinction spectra can be associated to absorption through plasmonic resonances.

The extinction peaks in this study are solely a result of the localized and collective plasmon resonances of the nanowire arrays. There are no Fabry-Perot effects observed in these samples. Additional Fabry-Perot resonant reflection (low transmittance) peaks are only observed in samples with nanowires above 700 nm in length. Longer nanowires result in stronger absorption peaks for our samples but do not change the location of the peaks because they are well below the 700 nm threshold.

At the ENZ wavelength, there exists field enhancement inside the metamaterial for the component of the electric field normal to the metamaterial air interface for ($p$)-polarized light. However, continuity of the displacement field ($\varepsilon_{air}(E_z^{inc} + E_z^{ref}) = \varepsilon_z E_z^{trans}$) dictates that this phenomenon can occur only if the dielectric constant with the ENZ occurs in the direction perpendicular to the interface ($\varepsilon_z \to 0$). Thus, ML metamaterials which have ENZ in the parallel component of the permittivity tensor ($\varepsilon_x \to 0$) do not exhibit the same field enhancement and absorption at ENZ wavelengths present in NW metamaterials. Furthermore, the NW ENP resonance is polarization insensitive and omnidirectional because the parallel component of the dielectric tensor ($\varepsilon_x$) interacts with both ($s$) and ($p$) polarized light. This makes ENP metamaterials attractive for thermal applications [2], [10]–[13], [24] and Salisbury screens [18] where omnidirectionality and polarization insensitivity is important. Our experimental work explains these effects using EMT.

Using EMT we can calculate the effective permittivities of the nanowire metamaterial which are described by the following equations:

$$\varepsilon_{xx} = \varepsilon_{yy} = \varepsilon_d \left[\frac{\varepsilon_m(1+\rho)+\varepsilon_d(1-\rho)}{\varepsilon_m(1-\rho)+\varepsilon_d(1+\rho)}\right], \quad \varepsilon_{zz} = \varepsilon_m \rho + \varepsilon_d(1-\rho) \quad (1)$$

Where $\varepsilon_d$ is the permittivity of the dielectric host, $\varepsilon_m$ is the permittivity of the metal nanowires and $\rho$ is the metal fill fraction ratio which is defined as:

$$\rho = \frac{nanowire\ area}{unit\ cell\ area} \quad (2)$$

For $\rho \ll 1$ the ENP is weakly dependent on fill fraction and the condition reduces to $\varepsilon_m + \varepsilon_d \approx 0$. On the other hand the ENZ resonance is strongly dependent on fill fraction and can be tuned by exploiting this fact.

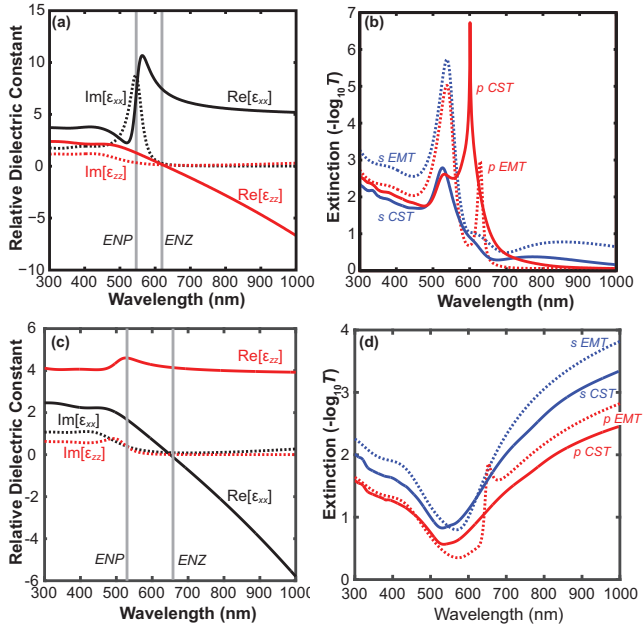

**Figure 2: Real and imaginary parts of Effective Medium Theory (EMT) permittivity in the parallel ($\varepsilon_{xx}$) and perpendicular ($\varepsilon_{zz}$) directions of a) Au nanowire array in alumina matrix metamaterials and c) Au-alumina multilayer (5 period) metamaterial with a fill fraction of 22% and 20% respectively. Simulated extinction spectra of b) Au nanowire array in alumina matrix metamaterial on glass substrate and d) Au-alumina multilayer (5 period) metamaterial from EMT and CST (full-wave simulation) under (s) and (p) polarization at 60 degrees plane wave incidence. Dashed blue line and dashed red line are extinction from EMT under s-polarization and p-polarization respectively. Solid blue line and solid red line are extinction from CST under s-polarization and p-polarization respectively. The parameters we use for a) and b) are as follows: radius 16nm, length 300nm and nanowire spacing 65nm, which gives a fill-fraction of 22%. The wavelength of T resonance and L resonance from CST is 524nm and 602nm respectively (523nm and 624 nm from EMT). The parameters we use for c) and d) are as follows: Au thickness 10nm, alumina thickness 40 nm, 5 periods, total thickness 250nm, which gives a fill fraction of 20%. The multilayer structure shows no absorption resonances unlike the nanowire design even though both designs possess epsilon-near-zero, epsilon-near-pole and hyperbolic response.**

Figure 2 a) shows the complex permittivity of a nanowire array with fill fraction of 22% [25]. A pole exists in $Re(\varepsilon_{xx})$ which dictates the location of the ENP resonance, while a zero exists in $Re(\varepsilon_{zz})$ dictating the location of the ENZ resonance. We confirm this through EMT and CST extinction simulations presented in Figure 2 b). ML metamaterial permittivities are shown in Figure 2 c) for a 5 period structure with a fill fraction of 20%. An ENZ is present in this plot as well as a much weaker ENP. The extinction spectrum of the ML structure simulated through EMT and CST is presented in Figure 2 d) where the absence of sharp absorption peaks (that are present in the NW extinction) can be noted.

## 3. Fabrication Procedure

Transparent NW metamaterial thin films are fabricated on clear rigid substrates. Glass slides are first cleaned with a 15 minute Piranha bath (3:1 sulfuric acid:hydrogen peroxide). A 20 nm $TiO_2$ layer is deposited as an adhesion layer between the glass substrate and a gold layer using atomic layer deposition (ALD). A subsequent 7 nm Au layer is deposited using DC magnetron sputtering, which will later act as a cathode during electrodeposition. Finally a 700 nm layer of Al is deposited on top of the Au layer using DC magnetron sputtering.

A two-step anodization procedure is performed on the Al layer to transform the Al metal into nano-porous $Al_2O_3$ called anodic aluminum oxide (AAO). Fortuitously, pores are formed via self-assembly in a quasi-hexagonal closed pack structure. The same electrolyte is used for each step [26] and this helps to improve pore order over a one-step procedure. The first anodization step uses 3% oxalic acid with a 30 V bias for 9 minutes. This initial $Al_2O_3$ oxide layer is removed with a wet etch in chromic phosphoric acid at 60°C for 2 hours. A second anodization step is performed again using 3% oxalic acid at ≈0°C with a 30 V bias until the entire aluminum layer is anodized through, making the sample transparent. This usually takes ≈14 minutes at which point the anodization current begins to rise and is allowed to reach ≈50 mA. The bias is subsequently decreased to 25 V, 20 V and 15 V for durations of 1 min each. This step-wise decrease in voltage helps to thin the oxide barrier layer that exists at the bottom of the pores after anodization, exposing the underlying Au cathode. To further thin the barrier layer, a wet etch in 5% phosphoric acid is performed for 20 minutes at room temperature. This wet etch removes any remaining barrier layer while slightly widening the pores, giving us control of nanowire diameter and fill fraction.

Au nanowires are deposited in the pores of the template using DC electrodeposition with a gold electrolyte (prepared using 0.05 M $HAuCl_4$, 0.42 M $Na_2SO_3$ and 0.42 M $Na_2S_2O_3$) and a bias of ∼ 1 V [3], [27]. Au from the electrolyte begins to plate the pore bottom as the exposed Au thin film acts as a cathode. Nanowire length is controlled by the duration of deposition and can range from tens of nanometers to the entire length of the pore in the AAO thin film. Deposition for our samples was carried out for a total of roughly 1-2 minutes resulting in a nanowire length of approximately 400 nm.

Depending on the AAO film thickness and the nanowires length, there may be some over or under deposition, leaving excess gold on the surface of the template or a void at the top of the pore. To level off the surface of the sample, an ion milling dry etch is performed. This leaves a surface where the Au nanowires are flush with the AAO thin film.

The samples are also annealed for 2 hours in an inert Ar atmosphere at 300°C [1]. This helps to increase the grain size of the metallic nanowires, allowing the electrons to have longer mean free path and reduced Drude damping. Figure 3 is a schematic that outlines the complete process flow. All steps are included in the schematic other than the annealing step as annealing does not alter the structure of the device.

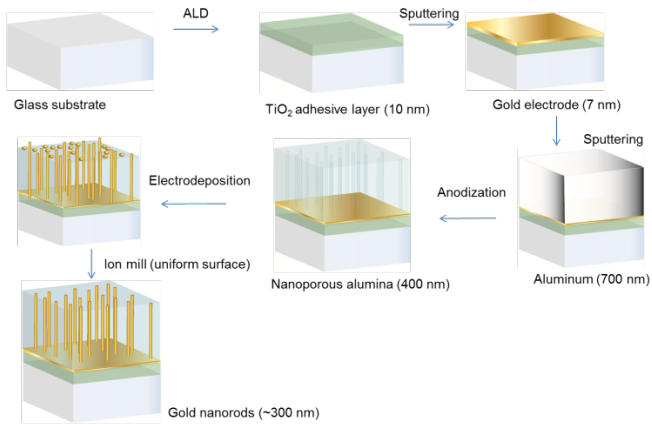

**Figure 3:** Schematic of process flow of device fabrication. Physical dimensions (diameter and length) and spacing of nanowires can be controlled by adjusting certain variables within the process flow. Our approach gives rise to large area uniform fabrication for the nanowire metamaterial.

### 4. Materials Characterization

To confirm the presence of Au nanowires SEM, EDX and XRD analysis are performed on the samples. Gold is chosen because it is an easy material to work with, it electrodeposits with a high yield of nanowires, does not oxidize easily and has a dielectric response that is ideal for plasmonics in the visible spectrum.

Figure 4 shows the pore ordering of the template surface. Using this image we can obtain values for pore diameter ($d$) and spacing ($S$) and calculate the metal fill fraction ($\rho$) using the following equation (assuming perfect hexagonal structure):

$$\rho = \frac{\pi (d)^2}{2\sqrt{3}(S)^2} \quad (3)$$

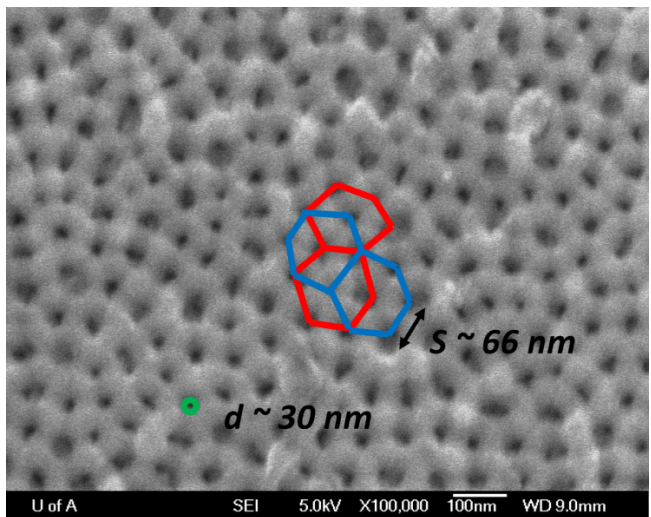

**Figure 4:** SEM image of AAO template quasi-hexagonal porous structure. Nanopore (and hence nanowire) diameter is approximately 30 nm, center to center pore spacing is approximately 66 nm. This produces a fill fraction of $\rho=19\%$. The template parameters (pore diameter, pore spacing, ordering, film thickness) can be altered by adjusting the anodization parameters.

A more rigorous method to extract the fill fraction from the above SEM image was also done to confirm an accurate fill fraction. The image is run through a code that outputs the ratio of "dark" pixels to "light" pixels. Given an accurate contrast threshold for the pore wall, "dark" pixels are defined as those pixels with a contrast below the boundary and represent the pore in the structure (where the metal is later deposited). Consequently, the "light" pixels are those with contrast above the threshold and represent the $Al_2O_3$ matrix.

Figure 5 shows the gold nanowire array after the alumina matrix has been removed using a 1 mM NaOH wet etch. This wet etch can cause the nanowires to clump together in groups due to the surface tension of the drying rinse water, as seen in the top-view SEM micrograph image shown in Figure 5 a).

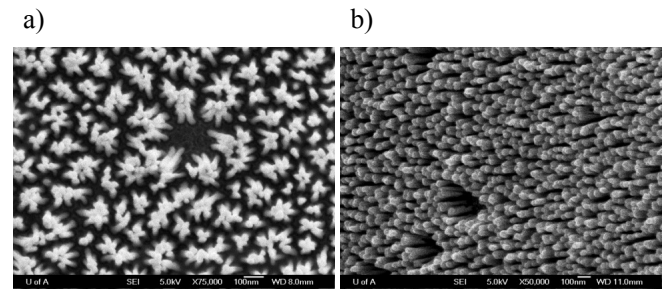

**Figure 5:** SEM images of a) top down and b) oblique views of Au nanowires grown in the pores of AAO templates. The AAO matrix has been removed in these images to view the nanowires, but is still present in the samples during transmission measurements. Because the template removal is done using a wet etch, the surface tension of the etchant/water pulls the nanowires together in to clumps as it dries. This is evident in a) as the nanowires can be seen leaning on each other in groups.

#### A. Energy Dispersive X-ray Spectroscopy

EDX analysis of an AAO template with gold nanowires deposited in the pores is compared with that of a template with empty pores to show which peaks are arising from the background elements present in the sample's substrate and matrix and which peaks can be attributed to the gold nanowires. There are several peaks visible in both sample spectra (Figure 6 a)) which correspond to the elements Si, Ca, Na, Mg, K (the glass substrate is predominantly Si with contaminants of Ca, Na, Mg, K), Ti (an adhesion layer of $TiO_2$ exists under the Au thin film) and Cr (a few nm of Cr is sputtered on top of all samples prior to SEM to ensure samples are suitable for SEM and to prevent surface charging artifacts in SEM images).

In Fig. 6 a), we notice peaks present only in the gold nanowire sample at 2.120 and at 9.712 keV. These are the two main characteristic peaks for gold which are a result of the M and $L_\alpha$ electron transmission lines respectively [28].

#### B. X-Ray Diffraction Analysis

To improve the optical properties of the gold nanowire arrays we investigated the effects of annealing on the crystallinity of the gold NWs. X-Ray Diffraction (XRD) was performed on samples prior to and after annealing. Gold peaks are observed in all samples for the (111), (200), (220), (311), (222) and (400) crystal planes (Figure 6 b)).

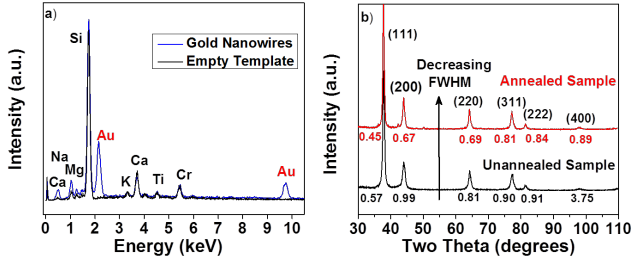

**Figure 6:** a) Energy dispersive X-ray analysis of empty template vs. gold nanowire filled template. Peaks corresponding to the elements Si (primary element in glass), Ca (glass impurity), Na (glass impurity), Mg (glass impurity), K (glass impurity), Ti (present in $TiO_2$ adhesion layer) and Cr (few nm conductive layer deposited on all samples for SEM imaging) are observed in both samples. Only in the gold nanowire sample spectrum are peaks observed at 2.1229 and 9.7133 keV, which correspond to the $M_{\alpha1}$ and $L_{\alpha1}$ electron transmission lines respectively, the characteristic lines of gold. b) X-ray diffraction pattern of gold nanowire samples before and after annealing. FWHM values are presented directly below each peak. Note that the FWHM value is larger prior to the annealing step for each individual peak.

Annealing has reduced the FWHM values of all Au plane peaks as shown in Figure 6 b). This suggests an increase in the overall grain size of the nanowires resulting in a larger mean free path of conduction electrons. A larger electron mean free path allows for lower optical losses. This ultimately leads to more efficient plasmon generation and higher quality factor resonances in the nanowires.

## 5. Experimental Results

Angular transmission spectroscopy is performed on several samples with fill fractions ranging from 10.5 - 26 % using an Ocean Optics fiber optic spectrometer and linearly polarized white light. Within the extinction spectra, two features can be observed in every sample (Figure 7). One absorption peak, located at ≈530 nm is attributed to the ENP resonance and another absorption peak, which is highly tunable in a range from 583-805 nm is attributed to the ENZ resonance.

These two absorption resonances are the result of localized plasmon polariton resonances supported by the NW structure. The ENP, also known as the transverse electric (TE) mode, is omnidirectional and its location is generally dependent on the materials used, although it is also very weakly dependent on the fill fraction. This is reflected in Figure 7 as well as Figure 8 b). Both show an ENP in the range of ≈530 ± 10 nm that is fixed in wavelength for any given angle of incident light. The TE mode is a plasmonic mode with resonant electron motion perpendicular to the nanowire axis and therefore is excited by both s and p polarized light [1].

The second absorption peak, located at the ENZ point, is known as the transverse magnetic (TM) mode. Contrary to the TE mode, the excitation efficiency of the TM mode is angularly dependent and highly sensitive to fill fraction as well as the materials used. Also, this resonance is only present in the absorption spectra using p-polarized light and does not exist for s-polarized light. The TM mode is a plasmonic mode with electron motion parallel to the nanowire axis [1], [29]. As the incident angle of light increases (measured from the normal as θ in Figure 7 a)), so does the amplitude of the absorption peak. The growing electric field vector parallel to the nanowire axis with increasing angle is the cause of this angularly dependent behavior. Using $Au/Al_2O_3$ nanowire arrays as has been fabricated in this report, the TM mode can be tuned from 583 nm (corresponding to a fill fraction of 26%) to 805 nm (corresponding to a fill fraction of 10.5%). Beyond the ENZ resonance wavelength the dielectric component $\varepsilon_{zz}$ becomes negative and the sample is hyperbolic. This forms the best approach to characterize the NW metamaterial with hyperbolic dispersion.

Scattering is one major source of uncertainty in extinction measurements as it is difficult to determine what fraction of the total power is being absorbed as opposed to scattered by the metamaterial. In an effort to remove scattering effects from our measurements, we have plotted $T_p/T_s$ (Figure 8 a) which allows scattering effects to be cancelled out as they are expected to affect s and p polarizations equally. We see a good agreement between CST simulations and our experimental data.

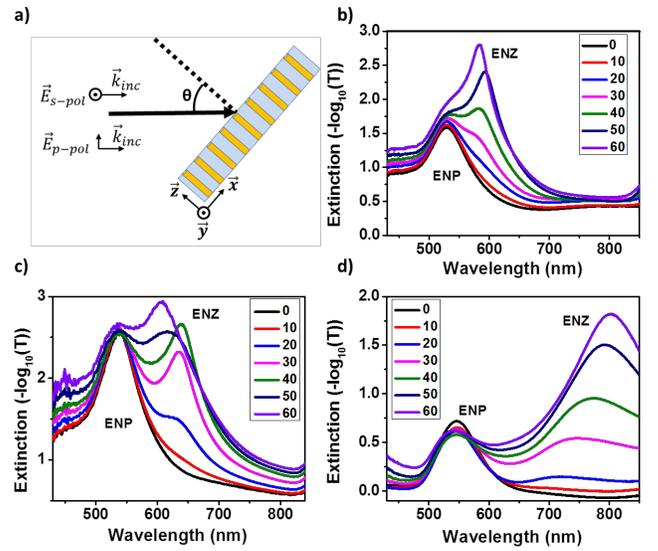

**Figure 7:** (a) Experimental setup showing how the extinction of our samples was measured. P-polarized light is defined as light with a maximum E-field perpendicular to the meta-surface (z direction in a) for a given angle while s-polarized light has its E-Field in the plane of the meta-surface (xy-plane in a) at all angles. (b)-(d) The experimental extinction spectra for p-polarized light of a gold nanowire system with fill fraction (b) 26%, (c) 23%, and (d) 10.5%. The ENP resonance is the shorter wavelength resonance for all three samples and occurs at $\lambda_{ENP} \approx 530 \pm 10$ nm. The ENZ extinction increases in intensity with an increase in incident angle from 0° to 60°. The location of the ENZ resonance shifts to longer wavelengths when we decrease the fill fraction from (b) $\lambda_{ENZ} \approx 583$ nm to (c) $\lambda_{ENZ} \approx 664$ nm to (d) $\lambda_{ENZ} \approx 805$ nm.

The ENZ and ENP points for the samples plotted in Figure 7 b)-d) are shown in Figure 8 b) along with the theoretical ENZ and ENP curves predicted by EMT (equations (1)). It is shown that the experimental and theoretical ENP resonances are very weakly dependent on the fill fraction values. Although it is stated earlier that this is only valid for $\rho \ll 1$, we can see that experimentally this holds true for much larger values of $\rho$. Alternatively, the ENZ point varies drastically along with the fill fraction and can be tuned within the range of 583-805 nm. Furthermore, within each ENZ point, a shifting of the ENZ resonance occurs as the angle of incident light is increased (larger symbols in Figure 8 b) corresponds to larger incident angles). The slight shift in peak wavelength of the plasmonic resonances is due to the interplay of competing effects of material loss and nonlocality. This will be addressed in future work. An overview of nonlocal effects can be found in reference [1]. Note the ENZ resonances are only present for p-polarized light. In stark contrast, the ENP resonance does not show within experimental uncertainty any spatial dispersion effects and has a fixed wavelength for varying angles of incident light.

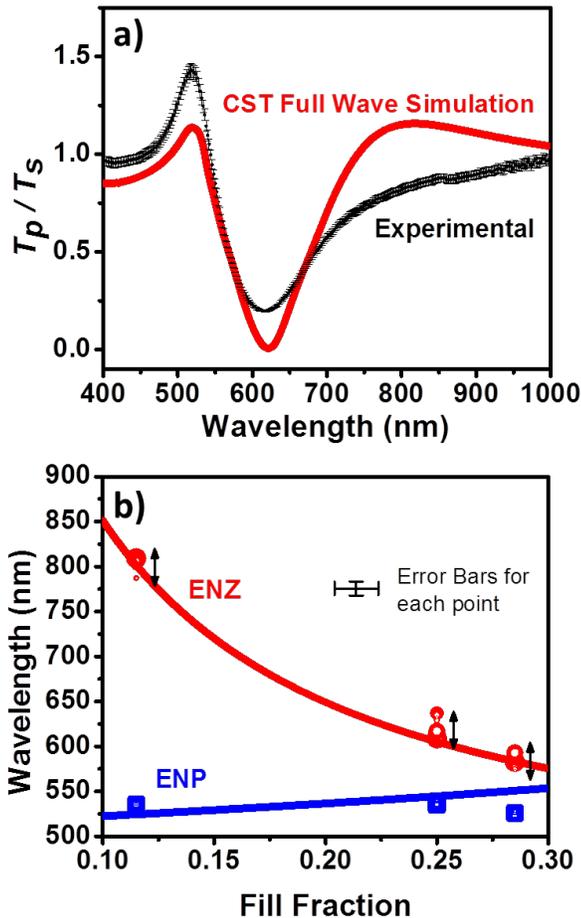

**Figure 8:** a) p-polarized transmittance divided by s-polarized transmittance from experiment and CST simulation at 30 degrees incidence. In experiment the surface roughness always leads to scattering thus usually lower transmittance. By comparing the transmittance ratio we can greatly reduce the scattering effect from optical measurement. The agreement between experiment and CST further confirm the experimental realization of nanowire metamaterial. b) Varying sized data points show experimental ENZ and ENP resonant wavelength's dependence on fill fraction and angle. The ENP resonance stays at a relatively fixed wavelength (≈530 ± 10 nm) for multiple fill fractions whereas the ENZ resonance can shift dramatically from ≈583 nm all the way to ≈805 nm for changing fill fractions. Data points are plotted for angles from 10˚-60˚ (symbol size is proportional to angle value) showing the spatial dispersion effect of the ENZ resonance. Solid lines show the theoretical location of the ENZ and ENP resonances as a function of fill fraction predicted using EMT (equations (1)).

## 6. Conclusion

We have contrasted the behavior of the omnidirectional and fixed wavelength ENP resonance with the angularly dependent and tunable ENZ resonance. While the ENP resonance remains at a wavelength of ≈540 nm in all samples, the ENZ resonance is highly dependent on the fill fraction of metal nanowire ($\rho$) and can be designed to exist in the range from ≈583-805 nm, corresponding to fill fractions of ≈26-10.5% respectively. We have distinguished the different structure and behavior of NW metamaterials from that of ML metamaterials and our experimental results agree with EMT and CST simulations. Future work will explore quantum, thermal and biosensing applications of such nanowire arrays.

**Funding.** Natural Sciences and Engineering Research Council of Canada (NSERC), Helmholtz Alberta Initiative.